\documentclass[12pt]{article}
\usepackage{a4}
\usepackage[latin1]{inputenc}
\usepackage{enumerate}
\usepackage{amssymb}
\usepackage{theorem}
\usepackage{exscale}
\usepackage{amsmath}
\usepackage[pagebackref,raiselinks,colorlinks,hyperindex]{hyperref}
\hypersetup{
    pdftitle={Wegner Estimate for Indefinite Anderson Potentials: Some Recent Results and Applications}, 
    pdfauthor={\textcopyright\ Vadim Kostrykin and Ivan Veseli\'{c} },
    pdfsubject={Wegner estimates for random Schrödinger operators with potentials of alloy type}, 
    pdfkeywords={Wegner estimate, random Schrödinger operators, multiscale analysis, localization, pure point spectrum}, }

\newcommand{\be}{\begin{equation}}
\newcommand{\ee}{\end{equation}}
\newcommand{\bea}{\begin{eqnarray*}}
\newcommand{\eea}{\end{eqnarray*}}
\newcommand{\beq}{\begin{eqnarray}}
\newcommand{\eeq}{\end{eqnarray}}
\newcommand{\nn}{\nonumber}
\newcommand{\RR}{\mathbb{R}}

\newcommand{\NN}{\mathbb{N}}
\newcommand{\ZZ}{\mathbb{Z}}

\newcommand{\EE}{\, \mathbb{E} \,}
\newcommand{\PP}{\mathbb{P}}

\newcommand{\NO}{| \| }
\newcommand{\ON}{\| | }

\newcommand{\g}{\ZZ^d}

\renewcommand{\H}{H_{\omega}}

\renewcommand{\r}{\right}
\renewcommand{\l}{\left}
\newcommand{\tL}{\mathop{\tilde\Lambda}}
\newcommand{\tT}{\tilde T}

\newcommand{\Lp}{\mathop{\Lambda^+}}
\newcommand{\cj}{\chi_j}
\newcommand{\la}{\langle}
\newcommand{\ra}{\rangle}
\renewcommand{\L}{\Lambda}
\renewcommand{\Lp}{\Lambda^+}
\newcommand{\Pe}{  {\cal P} \, (\eta )}

\newcommand{\vol}{\mathop{\mathrm{vol}}}

\newcommand{\supp}{\mathop{\mathrm{supp}}}
\newcommand{\Id}{{\mathop{\mathrm{Id}}}}
\newcommand{\Tr}{\mathop{\mathrm{Tr}}}

\newtheorem{thm}{Theorem}[section]

\newtheorem{prp}[thm]{Proposition}
\newtheorem{lem}[thm]{Lemma}

\newtheorem{asp}[thm]{Assumption}

\theorembodyfont{\normalfont}
\newtheorem{rem}[thm]{Remark}

\newenvironment{pro}{\pagebreak[2]  \mbox{} \\ {\bfseries Proof:} \par \nopagebreak }
            {\par \hspace*{\fill} {\bfseries q.e.d.} \\[1ex]}

\begin{document}

\title{Wegner Estimate for Indefinite Anderson Potentials: Some Recent Results and Applications}
\author{Vadim Kostrykin$\,{^1}$ and  Ivan Veseli\'{c}$\,{^2}$}
\date{}
\maketitle \vspace{0.3cm} {\noindent $^1$ {Fraunhofer-Institut
f\"{u}r Lasertechnik,} {Steinbachstra{\ss}e 15, D-52074 Aachen,
Germany}
{\ttfamily kostrykin@t-online.de, kostrykin@ilt.fhg.de}}\\[0.2cm]
{$^2$ {Fakult\"at f\"ur Mathematik,}
{Ruhr-Universit\"at Bochum, Germany}\\
{\ttfamily ivan@mathphys.ruhr-uni-bochum.de}, 
\href{http://www.ruhr-uni-bochum.de/mathphys/ivan}{\ttfamily www.ruhr-uni-bochum.de/mathphys/ivan}
}

\thispagestyle{myheadings}
\markright{\href{http://www.setsunan.ac.jp/mpg/confs/rims01/renom01.html}{``Renormalization Group Methods in Mathematical Sciences''}, \upshape RIMS, Kyoto, July 200}

\smallskip

{Appeared in \href{http://www.kurims.kyoto-u.ac.jp/~kyodo/kokyuroku/contents/1275.html}{volume 1275, (2002), 65--84.} of Lecture Notes of Research Institute for Mathematical Sciences, Kyoto University,}

\smallskip

\begin{abstract}
We review recent and give some new results on the spectral properties of  Schr\"odinger operators with a random potential of
alloy type. Our point of interest is the so called Wegner estimate in the case where the single site potentials change sign. The indefinitness of the single site potential poses certain difficulties for the proof of the Wegner estimate which are still not fully understood.

The Wegner estimate is a key ingredient in an existence proof of pure point spectrum of the considered random Schr\"odinger operators. Under certain assumptions on the considered models additionally the existence of the density of states can be proven.
\end{abstract}

\section{Introduction  and statement of results:
Alloy type models and Wegner's estimate}

The subject matter of this work are families of  Schr\"{o}dinger
operators $\{ H_\omega \}_{ \omega \in \Omega} $ acting on $L^2(\RR^d)$. They have been introduced as
quantum  mechanical models  for disordered media in solid state
physics. The random Schr\"odinger operator we consider is of {\em Anderson}
or {\em alloy} type and given by the following:

\begin{asp}[Alloy type model]
\label{atm}  Let
\begin{enumerate}[(i)]
\item
$V_0$ be a $\ZZ^d$-periodic potential, which is a infinitesimally
small perturbation of $-\Delta$ on $L^2(\RR^d)$, and $H_0 :=
-\Delta + V_0$ a periodic Schrödinger operator.
\item
$\omega :=\{\omega_k\}_{k \in \ZZ^d} \in (\Omega, \PP)$ be a
random vector composed of the coordinates $\omega_k$. Here $\Omega
= \times_{\ZZ^d} \RR, \PP:= \otimes_{\ZZ^d} \mu$, where $\mu$ is
the normalized Lebesgue measure on $[\omega_-, \omega_+]$.
\item
the {\em coupling constants} $\alpha_k  : \Omega \to \RR$ be
given by the projection $\alpha_k (\omega) := \omega_k, \, \forall
k \in \ZZ^d$. Then $\{ \alpha_k =\omega_k\}_{k \in \ZZ^d}$ forms
an iid sequence of random variables.
\item
the {\em single site potential}   $ u  $ be in $l^1(L^p)=\{f \in L^p_{\text{loc}}(\RR^d)| \ \|f\|_{l^1(L^p)} < \infty\}$ where
\[
\|f\|_{l^1(L^p)} := 
\sum_{k \in \ZZ^d} \bigg ( \int_{\|x\|_\infty < 1/2} |f(x-k)|^p dx \bigg )^{1/p}  
\]    
\item
the alloy type potential be given by the stochastic process 
\be
V_\omega(x) :=
\sum_{k \in \ZZ^d} \alpha_k(\omega) \, u (x-k)=
\sum_{k \in \ZZ^d} \omega_k \, u (x-k)  . 
\ee
\item
a family of Schr\"odinger operators be given by \be
H_\omega :=  H_0 + V_\omega,\ \omega \in \Omega . \ee
\end{enumerate}
\end{asp}
The above assumptions ensure by the Kato-Rellich theorem that each $H_\omega$ is a selfadjoint operator on the domain of the Laplacian.
\begin{asp}[Assumptions for the Wegner estimate]
\label{asp} Let additionally: 
\begin{enumerate}[(i)]
\item
\label{w} $\kappa >0$ and $\kappa \chi_{[0,1]^d} \le w \in
l^1(L^p(\RR^d))$, where $p:=p(d) = 2$ for $d \le 3$ and $p(d) >
d/2$ for $d \ge 4$.
\item
a partial ordering on $\RR^d \ni j,k$ be given by $j \succ k
\Leftrightarrow j_i \ge k_i \ \forall \ i = 1, \dots,d$.
\item
\label{a} $\Gamma \subset \{k \in \ZZ^d| \, k \succ 0\}$ be a
finite set, $a = \{a_k\}_{k \in \ZZ^d}$ be a so called {\em
convolution vector} with $a_k \neq 0 \Rightarrow k \in \Gamma$ and
$ a^* := \sum_{k \neq 0} |a_k| < a_0$.
\item
\label{step}
the single site potential be a {\em generalized step function}
\bea
 u (x) = \sum_{l\in \ZZ^d} a_l w(x-l).
\eea
\end{enumerate}
\end{asp}
For any cube $\Lambda_l=\Lambda_l(0)= [0,l[^d$ we can restrict
$H_\omega$ to $L^2(\Lambda_l)$ with appropriate boundary conditions
(b.c.). The results and proofs in this paper are equally valid if
we chose for the restriction $H_\omega^l$ Dirichlet, Neumann  or
periodic b.c. We denote the spectral projection of $H_\omega^l$ on
the energy interval $I = ]E_1, E_2[$ by $P_\omega^l (I)$ and the
characteristic function of the unit cube $ \Lambda_1(j)= [0,1[^d +
j$ at the lattice site $j \in \ZZ^d$ by $ \chi_j$. The
expectation w.r.t.~$\PP$ is denoted by $\EE$. Our Wegner estimate \cite{Wegner-81} reads:

\begin{thm}
\label{theorem1} For all $E_2 \in \RR$ there exist a constant $C=
C(E_2)$ such that for all $l \in \NN$ and $E_1 \le E_2$  we have
\be
\label{resultat1}
\EE \l [ \Tr P_\omega^l (]E_1,E_2[) \r ] \le C \,
(\omega_+ -\omega_-)^{-1} \, (E_2-E_1) \, l^d .
\ee
\end{thm}

\begin{rem}
\label{anterem} By replacing the convolution vector $a$ with $\kappa a$ we may assume
$\kappa =1$ in Assumption \ref{atm} (\ref{w}). Furthermore, by rescaling the support of
$\mu$ we may assume $a_0 =1$. Note that by adding a part of the periodic potential to
$V_\omega$ we can assume without loss of generality that the support of $\mu$ starts at
$0$, i.e. $ \supp \mu = [0, \omega_+] $ for some $ \omega_+ >0 $.
Our results are also true, if we have $a_0 = -1$ and $a^*<1$ in
our model. In this case, in the proofs everywhere where positivity
is used, negativity has to be used instead.
\end{rem}

In the next section we deduce the existence of the density of states from the Wegner estimate in Theorem \ref{theorem1} and discuss its role for the proof of localization.
Section 3   contains the proof of  Theorem \ref{theorem1}
and Section 4 reviews earlier results for indefinite alloy type models.
\vspace{1em}

{\bfseries Acknowledgements:}

The second named author is grateful for stimulating discussions with N.~Ueki and K.~Veseli\'c, he
thanks the the Japanese Society for the Promotion of Science, the Research Institute for Mathematical Sciences, Kyoto University for financial support and K.~Ito, S.-I.~Kotani and N.~Minami for hospitality at the RIMS  and the Universities of Osaka and Tsukuba.

\section{Density of states and  localization}
Under our assumptions the family $H_\omega, \omega \in \Omega$ fits into the
general theory of ergodic random Schr\"odinger operators
\cite{Kirsch-89a,CarmonaL-90,PasturF-92}. We infer two central results
from this theory.
\begin{enumerate}[(A)]
\item
\label{genthm1}
The spectrum of the family $H_\omega, \omega \in \Omega$ is non-random in the
following sense. There exists a subset $\Sigma$ of the real line and an
$\Omega' \subset \Omega$, $ \PP (\Omega') =1 $ such that for all
$ \omega \in \Omega' $ one has
$
\sigma (H_\omega) = \Sigma
$.
The analogous statement holds true for the essential, discrete,
continuous, absolutely continuous, singular continuous, and pure
point part of the spectrum. Note that the pure point spectrum
$\sigma_{pp}$ is the closure of the set of eigenvalues of $H_\omega$.
\item
\label{genthm2}
There exists a {\em self averaging} integrated density of states
 associated with the family $H_\omega, \omega \in \Omega$.
This means that the normalized eigenvalue counting functions
\begin{equation}
N_\omega^l (E) = l^{-d} \# \{ i | \ \lambda_i (H_\omega^l) < E \}
= l^{-d} \Tr P_\omega^l(]-\infty,E[)
\end{equation}
of $H_\omega^l$ converge for almost all $\omega$ to a limit
$N := \lim_{ l \to \infty} N_\omega^l $ which is $\omega$-independent. For definiteness we use periodic b.c.~in the construction of $H_\omega^l$.
\end{enumerate}

We call $N$ the {\em integrated density of states} (IDS) of $H_\omega$ and
$N_\omega^l$ the {\em finite volume} IDS of $H_\omega^l$.

\begin{rem}
\label{questions}
While the two above facts (\ref{genthm1}) and (\ref{genthm2}) follow from the
general theory, one is interested in more detailed spectral properties of
specific models $H_\omega, \omega \in \Omega$, e.g.:
\begin{itemize}
\item
Which spectral types can occur in $ \sigma ( H_\omega) $?
\item
Can something be said about the regularity of the IDS $N$ as a
function of the energy $E$? Is it H\"older continuous or does even its derivative, the {\em density of states} exist.
\end{itemize}
\end{rem}

Our result on the regularity of the IDS is strong enough to imply the existence of the density of states:

\begin{thm}[Density of states]
\label{theorem2}
Under the assumptions of Theorem \ref{theorem1} the IDS of the alloy type model
 $\{ H_\omega \}_{ \omega \in \Omega} $ is Lipschitz continuous: for all $E \in \RR$ there exists a constant $C$ such that
 \begin{equation}
N(E) - N(E-\epsilon) \le C \, \epsilon , \quad \forall \ \epsilon \ge 0
\ .
\end{equation}
It follows that the derivative $\frac{dN}{dE}$ exists
for almost all $E$.
\end{thm}
\begin{rem}
The theorem  follows directly from (\ref{resultat1}) and the self averaging property $N (\cdot) =\EE N(\cdot)$.
\end{rem}

The second question of Remark \ref{questions} is related to the  transport properties of
the medium modelled by $H_\omega$. A perfect crystal is described by a
Schr\"odinger operator with periodic potential. It has purely absolutely
continuous spectrum, which reflects its good electric transport
properties. In contrast to this, it has been proven that random
perturbations of this regular structure give rise to energy intervals
with pure point spectrum. This corresponds to the less effective
transport properties of random media. The existence of pure point
spectrum in this context is called {\em localization}.

Now we  indicate the general scheme of the proof of localization and where the Wegner estimate enters.
An intermediary step in the proof of localization is the establishing of the exponential
decay of the resolvent
\be
\label{ExpDecOnR}
\sup_{ \epsilon \neq 0} \| \chi_{x} R(\epsilon) \chi_y \|_{ {\cal L} ( L^2 (\RR^d))}
\le  const \, e^{ -c |x-y|}
\mbox{  for almost all } \omega  \ ,
\ee
where $ R(\epsilon) :=  ( H_\omega -E -i \epsilon )^{-1}$ is the resolvent of $ H_\omega $
near an energy value $ E $ in the energy interval $ I  \subset \RR$
(typically near a boundary of $\sigma(H_\omega)$) for which we want to prove localization.
The $ \chi_x $ and $ \chi_y $  are characteristic functions of unit cubes
centered at $ x $, respectively at $ y $. This bound can be used to rule out absolutely
continuous spectrum \cite{MartinelliS-85} and is interpreted as absence of diffusion
\cite{FroehlichS-83,MartinelliH-84} in the energy
region $ I $  if  (\ref{ExpDecOnR}) holds for all $ E \in I $.

It turns out that the finite size resolvent $ R_\Lambda(\epsilon) :=
( H_\omega^\Lambda - E -i \epsilon )^{-1} $ is easier approachable
than $ R(\epsilon) $ on the whole space.
Here $ H_\omega^\Lambda $ is the restriction
of $ H_\omega $ to $ L^2 ( \Lambda ) $ with some appropriate boundary conditions;
the use of Dirichlet or periodic b.c.~is most common.
However the operator $ H_\omega^\Lambda $ is not ergodic and for its resolvent an estimate
like (\ref{ExpDecOnR}) can be expected to hold only with a probability
strictly smaller  than one.
This is the place where multi scale analysis (MSA) enters. It is an
induction argument over increasing length scales $ l_j $. They are defined recursively
by $ l_{j+1} := [ l_j^\zeta]_{3} $, where $ [ l_j^\zeta]_{3} $ is the greatest multiple of $3$ smaller than
$ l_j^\zeta $. The scaling exponent $ \zeta $ has to be from the interval $ ]1,2[$.
On each scale one considers the box resolvent
$ R_j(\epsilon) := R_{\Lambda_{l_j}}(\epsilon) $ and proves  its exponential decay with a
probability which tends to $1$ as $ j \to \infty $. We outline briefly
the ingredients of the MSA as it is given in \cite{CombesH-94b,KirschSS-1998a}
or \cite{CarmonaL-90}.

First we explain some notation which is used afterwards. Let  $ \delta > 0 $
be a small constant independent of the length scale $ l_j$ and $ \phi_j(x) \in C^2$
a function which is identically equal  to $0$ for $x$ with $ \| x\|_\infty > l_j -\delta$
and identically equal to one for $x$ with $ \| x\|_\infty < l_j - 2 \delta $.
The commutator
$ W(\phi_j ) := [-\Delta, \phi_j]:= -(\Delta \phi_j) -2  (\nabla \phi_j) \nabla$
is a local operator acting on functions which live on a ring of width $ \delta $
near the boundary of $ \Lambda_j := \Lambda_{l_j} $. We say that a pair
$ ( \omega,\Lambda_j ) \in \Omega \times {\cal B } ( \RR^d) $ is {\it
$m$-regular}, if
\be
\label{ExpDecOnLj}
\sup_{\epsilon \neq 0} \| W (\phi_j ) R_j( \epsilon ) \chi_{l_j/3} \|_{\cal L}
\le e^{-ml_j} \ .
\ee
Here $\| \cdot \|_{\cal L}$ is the operator norm on $L^2(\Lambda_j)$ and
$ \chi_{l_j/3} $ the characteristic function of
$ \Lambda_{l_j/3} := \{ y| \, \| y  \|_\infty \le l_j /6 \} $. Thus the
distance of the  supports of $ \nabla \phi_j $ and $ \chi_{l/3} $ is at least
$ l_j/3 -2\delta \ge l_j/4 $.

Let $q_0 > 0$ and $m_0 \ge const \, l_0^{-1/4}$. The starting point of the MSA is the estimate
\bea
\mbox{(H1)} (l_0,m_0, q_0 ) \makebox[12ex]{}
\PP \{ \omega | \, ( \omega , \Lambda_0 ) \mbox{ is $m_0$-regular} \} \ge 1 - l_0^{q_0}
\eea
which serves as the base clause of the induction. The induction step consists in proving
\be
\label{H1Step}
\mbox{(H1)} (l_j,m_j, q_j ) \Longrightarrow  \; \mbox{(H1)} (l_{j+1},m_{j+1}, q_{j+1} )
\ee
For the mass of decay $ m_{j+1} $ and the probability exponent
$ q_{j+1} $ on the scale $ l_{j+1} $ the following estimates are valid
\beq
\lefteqn{\forall \xi > 0  \ \exists c_1, c_2 , c_3 \mbox{  independent of $j$ such that}}
\nn
\\
m_{j+1}
& \ge &
 m_j \l ( 1 - \frac{ 4 l_j}{ l_{j+1}} \r ) - \frac{c_1}{l_j} -c_2 \frac{\log l_{j+1}}{l_{j+1}}
\label{mj+1}
\\
\label{qj+1}
l_{j+1}^{q_{j+1}}
&  \le &
 c_3 \l ( \frac{l_{j+1}}{ l_j} \r)^{2d} l_{j}^{2q_j} + \frac{1}{2} l_{j+1}^{-\xi} \ .
\eeq
For the recursion clause (\ref{H1Step}) a Wegner estimate as in \eqref{resultat1}   is needed:
\bea
\mbox{(H2)}  \makebox[12ex]{}
\PP \{ \omega | \, d( \sigma (H_\omega^\Lambda ), E ) \le \epsilon \}
\le C_W  \: \epsilon | \Lambda |^2
\eea
for all boxes $ \Lambda \subset \RR^d $ and all $ \epsilon > 0 $,  such that $ [E-\epsilon,E+\epsilon] $
is contained in neighbourhood of $I$.
Here $ | \Lambda | $ stands for the Lebesgue measure of the cube $ \Lambda $.

The deterministic part of the induction step uses the {\it geometric resolvent formula}
\cite{CombesH-94b,HislopS-96}
\be
\phi_\Lambda ( H_{\Lambda '} - z )^{-1} = ( H_{\Lambda} - z )^{-1} \phi_\Lambda +
( H_{\Lambda} - z )^{-1} W (\phi_\Lambda) ( H_{\Lambda '} - z )^{-1}
\ee
for $ z \in \rho (H_{\Lambda '}) \cap  \rho ( H_\Lambda ) $ and $ \phi_\Lambda \in C^2$ with support in
$ \Lambda \subset  \Lambda ' $. It gives the estimate
\be
\| \chi_{l/3}(\cdot - x) R_{3l'} (\epsilon)   \chi_{l/3}(\cdot-y) \|_{\cal L}
\le ( 3^d e^{-ml})^{3|x-y| l^{-1} -4 } \| R_{3l'}(\epsilon) \|_{\cal L}
\ee
if no two disjoint non-regular boxes $ \Lambda_l \subset \Lambda_{l'} $ with center in
$ \frac{l}{3} \g \cap \Lambda_{3l'} $ exist for $ \omega $.
In our case $ l := l_j$ is the length scale on which the exponential decay of the resolvent is already known and $ l' := l_{j+1} $ the scale on which we want to prove it.
By the estimates (H1),(H2) we have with probability $1-l_{j+1}^{q_{j+1}} $ (bounded by the inequality (\ref{qj+1})) exponential decay on the length scale
$l_{j+1}$ with mass $m_{j+1}$ (bounded as in (\ref{mj+1})).

We stated above the ingredients of the MSA as they are valid if $ u $ is compactly supported.
If the single site potential is of long range type (as in (\ref{MinDecOfu}) below)
one has to use the adapted MSA from the papers \cite{KirschSS-1998a,Zenk-1999}.

Once the estimate (H1) is established on all length scales $ l_j, j \in \NN $, one infers
an exponential decay estimate for the resolvent on the whole of $ \RR^d$. Afterwards one
uses a
spectral averaging technique (cf.\cite{CombesH-94b}) based on ideas of Kotani, Simon, Wolf and
Howland to conclude localization \cite{KotaniS-87,SimonW-86,Howland-87a}.
An alternative version of the MSA can be found in the monograph \cite{Stollmann-2001} (see also \cite{GerminetK-2001a,GerminetK-2001b}.

Recent papers concentrate on proofs for the Wegner estimate and the initial length scale
decay of the resolvent. At the same time adaptations of the MSA for various random
Schr\"odinger operators, as well as Hamiltonians governing the motion in classical physics
appeared \cite{FigotinK-1996,FigotinK-1997a,CombesHT-1999,Stollmann-1998}.

We discuss briefly some results for quantum mechanical Hamiltonians.
For $V_\omega $ a Gaussian random field a Wegner estimate
was shown in \cite{FischerHLM-1997}. Its main feature is that no underlying lattice structure of
 $ V_\omega $ is needed.  This result allows one to conclude localization  for the
corresponding Schr\"odinger operator at low energies \cite{FischerLM-2000}. Kirsch, Stollmann and
Stolz proved in \cite{KirschSS-1998a} (cf.~also \cite{Zenk-1999}) a Wegner estimate with only polynomial decay conditions on the
single site potential $u$ and deduced a localization result for Hamiltonians
with long range interactions. They require
\be
\label{MinDecOfu}
|u(x)| \le const \,  (|x|+1 )^{-m}  \mbox{ for some } m > 4d  \ .
\ee

 The resolvent decay estimate (H1) for some initial length scale can be proved
with semiclassical techniques. Using the Agmon metric one can achieve rigorously
decay bounds with what is called among physicists WKB-method \cite{CombesH-94b,HislopS-96}.
However this reasoning is only applicable for energies near the bottom of the spectrum.

The so-called {\it Combes-Thomas argument} \cite{CombesT-73} allows one to infer the following inequality
\be
\label{CoThBound}
\| \chi_x (H-z)^{-1} \chi_y \|_{\cal L} \le \big[const \,  d(\sigma(H),z)\big]^{-1}  \,
e^{-const \, d(\sigma(H),z) \, |x-y| }
\ee
where $H$ is a self-adjoint Schr\"odinger operator on $ L^2 (\RR^d) $ and $ z \in \rho(H) $.
It was first applied to multiparticle Hamiltonians \cite{CombesT-73}, but it is also useful in
our case, as soon as we get a lower bound on  $ d(\sigma(H_\omega^\Lambda),z) $.
Thus it is  sufficient to prove an estimate like
\be
\label{ProbNoSpecInI}
\PP \{ \omega | \ d(\sigma( H_\omega^l ,I)  < l^{- \alpha } /2  \: \} \le  l^{-q}
\ee
for some $ \alpha \in ]0, 1/4] $. Now Inequality (\ref{CoThBound})
implies the initial scale estimate (H1) with $ m_0 \ge const \, l^{-1/4} $ for $l$ large and $E\in I$, cf.~
\cite[Lemma 5.5]{KirschSS-1998a}.
The constant depends on the energy and the potential, but not on $l $ and $m_0$.

Two possibilities were used to deduce (\ref{ProbNoSpecInI}). The first is to assume a special disorder regime,
more precisely to demand a sufficiently fast decay of the density $g$ of the distribution of $ \omega $
near the endpoints $ 0$ and $ \omega_+ $ of $ \supp g$:
\bea
\lefteqn{\exists \tau > d/2 \ : \ \forall \mbox{ small  } \epsilon >0  } & &
\\
& &
\int_0^\epsilon g (s) ds \le \epsilon^\tau , \mbox{ respectively }
\int_{\omega_+ -\epsilon }^{\omega_+ } g(s) ds \le \epsilon^\tau
\eea
depending on whether one wants to consider an energy interval $I$ at a lower or upper spectral edge. This approach
was used in \cite{CombesH-94b,KirschSS-1998a}. Its shortcoming is that it excludes quite a few
distributions, e.g. the uniform distribution on $[0,\omega_+]$.

The other way to prove (\ref{ProbNoSpecInI}),  is to use the existence
of Lifshitz tails of the integrated density of states at the edges of the spectrum:
One can show that for a variety of types of random Schrödinger operators, including ours, the IDS does not change,  if one replaces the periodic b.c.~in its definition by Dirichlet b.c.:
\beq
\label{DefIDS}
N(E) =
\lim_{\Lambda \nearrow \RR^d}  |\Lambda|^{-1}
\# \{ \mbox{ eigenvalues of } H_\omega^{\Lambda,D}
\mbox{ below } E \} \ ,
\eeq
i.e. one considers the IDS as the limit of the normalized counting function of eigenvalues of the Dirichlet Hamiltonian $ H_\omega^{\Lambda,D} $ on  $ L^2 (\Lambda) $. 
The use of Dirichlet b.c.~in in the above formula for the IDS implies \cite{KirschM-82c}
\be
\label{SupIDS}
N(E) = \sup_{\Lambda \nearrow \RR^d} N( H_\omega^{\Lambda,D} ,E ) \ .
\ee
One says that $ N( \cdot ) $ exhibits {\em Lifshitz tails} at some spectral edge $ \cal E $ if
\be
\label{LifTaildef}
\lim_{E \to \cal E} \frac{\log |\log | N(E) - N(\cal E) | |}{\log |E -\cal E|}
= -\frac{d}{2}   \  .
\ee
At the infimum of the spectrum, i.e. for $ {\cal E} = \inf \sigma ( \H ) $, (\ref{SupIDS})
and (\ref{LifTaildef}) imply
\bea
\# \{ \mbox{eigenvalues of } H_\omega^{\Lambda,D} \mbox{ in } [{\cal E},E] \}
\le | \Lambda | N(E) \le | \Lambda | \exp ( -c E^{-d/4} )
\eea
since $N({\cal E} ) = 0$.
This estimate was used in \cite{Klopp-95a} together with a \v{C}ebi\v{s}ev inequality
to prove (H1) at the bottom of the spectrum, see also \cite{MartinelliH-84}.
For internal spectral edges the situation is similar, however one needs to know some additional properties of the unperturbed periodic operator $H_0 =-\Delta +V_0$, see \cite{Klopp-1999,Veselic-1998}.

If one considers the situation where the single site potential changes sign the initial scale estimate has been established only under restrictive hypotheses \cite{Veselic-2000b,HislopK-2001}.

\section{Proof of Theorem \ref{theorem1}}
\label{proof}

Let $\tilde{\Lambda} := \Lambda \cap \ZZ^d$ be the lattice points
in $\Lambda= \Lambda_l$. As in  \cite{CombesH-94b} we estimate
\begin{equation}
\label{CH} \EE \l [ \Tr P_\omega^l(I) \r ] \le e^{E_2} C_V \sum_{j
\in \tilde{\Lambda}} \, \l \|  \EE \l [ \chi_j P_\omega^l (I)
\chi_j \r ] \r \| .
\end{equation}
where the constant $C_V$ is an uniform upper bound on $\Tr (\chi_j
e^{-H_\omega^{\Lambda+j}} \chi_j)$, cf.~proof of Theorem 76 in
\cite{ReedS-78}. Thus for the proof of Theorem \ref{theorem1} it is
sufficient to prove the following proposition dealing with the
expectation of a quadratic form.

\begin{prp}
\label{tech} Let $\Lambda = \Lambda_l$ for some $l \in \NN$. For $
f \in L^2(\Lambda_l)$ there exists a constant
$ C$ such that for all $j \in \tL$
\be \label{qf} \EE \la f , \cj P_\bullet^l (I) \cj
f \ra \le C \, \omega_+^{-1} \, |I| \, \|f\|^2  .
\ee
\end{prp}

\begin{pro} It suffices to consider the case $ \|f\| =1 $.
Assume first $ w = \chi_0$. Denote by  $\L^+$ the set  $ \tL - \Gamma := \{ k - \gamma |
\, k \in \tL, \gamma \in \Gamma \}$ of lattice sites in $\ZZ^d$ which influence the value
of the potential in the cube  $\L$ and by $L = \# \Lp$ its cardinality. The convolution
vector $a$ defines a (block) Toeplitz matrix $A := \{ A_{j,k} \}_{j,k\in \Lp}, \ A_{j,k}:=
a_{j-k}, \forall \, j,k\in \Lp $. Note that the coupling constants with index outside
$\Lp$ do not influence the random variable $P_\omega^l$ in (\ref{qf}). So we may pass on
to a "smaller" probability space $\Omega = \RR^L$ and consider the linear transformation
$ A : \RR^L \to \RR^L$, $A \omega= \eta$ for vectors $ \omega  :=\{\omega_k\}_{k \in
\Lp}$ and $ \eta  :=\{\eta_k\}_{k \in \Lp}$. By Assumption \ref{atm} (\ref{a}) the
inverse $B$ of $A$ exists and has its column sum  norm $\NO B \ON_1 $ bounded by
$\frac{1}{1-a^*}$, cf.~\cite[Sec.~4.4]{Veselic-2001}.

The random variable $ \omega_0$ has the density $ g(x) =
\frac{1}{\omega_+} \chi_{[0,\omega_+]} (x)$. Thus $G(\omega) :=
\prod_{j \in \Lp} g(\omega_j)$ is the common density of $\omega$
and $K(\eta): = | \det B| \, G(B \eta)$  the one of $\eta$.

We calculate the representation of the alloy type potential in the
new coordinates $\eta$. For $ x \in \Lambda$ \be V_{ B \eta } (x)
=V_\omega(x) = \sum_{k \in \Lp} \omega_k \sum_{l \in \Gamma} a_l
\chi_{k+l} (x) = \sum_{j \in \tL} \eta_j \cj(x) . \ee This
representation particularly shows that for any fixed $j \in \tL$
we have a one parameter family of  potentials,
cf.~\cite{FischerHLM-1997} \be \eta_j \mapsto \l ( \sum_{j \neq k
\in \tL} \eta_k \chi_k \r ) + \eta_j \cj \ee which is linearly
increasing locally on $\Lambda_1(j)$. This fact will later enable
us to apply results from \cite[Sec.~4]{CombesH-94b}. Using the
abbreviation
    \begin{equation}
     {\cal P} \, (\eta ) := \la f, \chi_j  P_{B\eta}^l (I) \chi_j f \ra .
    \end{equation}
the integral transformation of (\ref{qf}) reads \be \label{int}
\EE \la f , \cj P_\bullet^l (I) \cj f \ra =\int_{\RR^L} d \eta \,
k (\eta) \, \la f , \cj P_{B \eta}^l (I) \cj f \ra = \int_{\RR^L}
d \eta \, k (\eta) \,    {\cal P} \, (\eta ) . \ee The integration
domain $M := A ( [0,\omega_+]^L)$ in (\ref{int}) is a compact set,
thus for $t >0$ \be \label{sup} (\ref{int}) \le \sup_{\eta \in M}
\l [ k(\eta) (1 + t \eta_j^2) \r ] \int_{M} d \eta \,  \frac{
{\cal P} \, (\eta ) }{1 + t \eta_j^2}  . \ee The achievement of
the last inequality is that we introduced an artificial density $
\frac{1}{1 + t \eta_j^2}$ with which we can deal better
analytically and, more important, that we decoupled the dependence
of the density on $\eta_j$ and on the  other components of $\eta$.
Now \be \sup_{\eta \in M} \l [ k(\eta) (1 + t \eta_j^2) \r ] \le
|\det B | \, \omega_+^{-L} \, ( 1 + t \NO A \ON_1^2 \eta_j^2) \ee
leaves us with the analysis of the integral on the rhs of
(\ref{sup}). In the next step we will decouple the dependence of
the integration domain $M$ on $\eta_j$ from the dependence on the
other components of $\eta$. For this aim we will factorize $M$
similarly as in \cite[Lem.~4.5.11]{Veselic-2001}.

Lemma \ref{trilem} below tells us that $B$ inherits from  $A$ the
triangular property \be \label{3angle} B_{kk} = 1 \text{ and }
B_{lk} \neq 0 \Rightarrow l \succ k, \quad \forall \, l,k \in \Lp.
\ee For a pair $l,k \in \ZZ^d$ which does not satisfy $l \succ k$
let us write $l \not \succ k$. We will need the following
decomposition of $\Lp$ and $\eta$ adapted to the lattice site $j
\in \ZZ^d$.
\begin{align}
\Lp = \L_< \cup \{ j \} \cup \L_>, \quad & \L_< = \{ k \in | \, k
\not \succ j \}, \ & \L_> = \{ n \in | \, n  \succ j, n \neq j \}
\\
\eta = ( \eta_< , \eta_j, \eta_>), \quad & \eta_< = \{ \eta_k | \,
k \in \L_< \}, \ & \eta_> = \{ \eta_n | \, n \in \L_> \} .
\end{align}
Then:
\begin{align}
M & = \{ \eta | \,  B\eta \in [0,\omega_+]^L \}
\\
  & = \Bigg \{ \eta \Bigg |
  \begin{array}{rllr}
  \sum_{l \in \L_<} B_{kl} \eta_l & \in [0, \omega_+] &                                  & \forall \,  k \in \L_<
  \\
  \eta_j                          & \in [0, \omega_+] & -\sum_{l \in \L_<} B_{jl} \eta_l
  \\
  \sum_{l \in \L_>} B_{nl} \eta_l & \in [0, \omega_+] & -\sum_{l \in \L_<} B_{nj} \eta_j - \sum_{l \in \L_<} B_{nl} \eta_l & \forall \, k \in \L_>
  \end{array}
  \Bigg \}
\nn
\end{align}
Set now \bea \xi  = \xi (\eta_>) = -\sum_{l \in \L_<} B_{j,l}
\eta_l \ \text{ and } \  \Xi = \Xi (\eta_<) = - \sum_{l \in \L_<}
b_{l} \eta_l \eea
where $ b_l := \{ B_{nl} \}_{n \in \L_> } $  is a column vector,  and
\begin{align}
M_< &  := \l \{ \eta_< | \, \sum_{l \in \L_<} B_{kl} \eta_l  \in [0,
\omega_+]  \forall \,  k \in \L_<  \r \}
\\
\nn M_j ( \eta_< ) &  := \{ \eta_j | \,  \eta_j \in [0,\omega] +
\xi \}  =  [\xi , \xi \omega_+ ]
\\
\nn M_> (\eta_<,\eta_j)& := \l \{ \eta_> | \,   \sum_{l \in \L_>}
B_{nl} \eta_l  \in [0, \omega_+]  -\sum_{l \in \L_<} B_{nj} \eta_j
- \sum_{l \in \L_<} B_{nl} \eta_l  \forall \, k \in \L_>  \r \} .
\end{align}
Write the integral in (\ref{sup}) as:
\[
\int_{M_<} d\eta_< \int_{M_j(\eta_<)} d\eta_j
\int_{M_>(\eta_<,\eta_j)} d \eta_>  \, \frac{\Pe}{1 + t \eta_j^2}
.
\]
Note that we can write the integral in this ''successive'' form
only because property (\ref{3angle}) holds.

We would like to apply the spectral averaging result of
\cite[Section 4]{CombesH-94b} to the integral $\int_{M_j(\eta_<)}
d\eta_j $. The integration over $\eta_<$ causes no problem because
it stands outside the $d \eta_j$-integral. However, the
integration domain $M_> (\eta_<,\eta_j)$ of the ''inner'' integral
is a function of $\eta_j$, so we cannot pull this integral out of
$\int_{M_j(\eta_<)} d\eta_j $. To solve this problem we will
carefully enlarge the domain $M_>(\eta_<,\eta_j)$ so that it
becomes  $\eta_j$-independent. In doing so we have to make sure
that the enlargement is not too ''generous''. More precisely, the
factor by which the volume of the domain increases has to  remain
bounded as $\L$ tends to $\RR^d$. If one enlarges $M_>(\eta_<.
\eta_j)$ too naively one can incur a factor growing exponentially
in $L = \# \Lp$, cf.~\cite[Remark 4.5.8.]{Veselic-2001}.

Fix $\eta_< \in M_<$ and thus  $\xi $ and $\Xi$. Now $M_> (\eta_<,
\eta_j)$ is for all values of $\eta_j \in [\xi, \xi +\omega_+]$
contained in
\[
M_>^+ (\eta_<) := \bigcup_{s \in  [\xi,\xi + \omega_+]}
\l \{ \eta_> | \, \sum_{l \in \L_>} B_{nl} \eta_l \in [0,\omega_+] - s B_{nj} +
\Xi_n , \ n \in \L_> \r \} .
\]
Thus
\[
\int_{M_j(\eta_<)} d\eta_j \int_{M_>(\eta_<,\eta_j)} d \eta_>
\frac{\Pe}{1 + t \eta_j^2} \le \int_{M_>^+(\eta_<)} d \eta_>
\int_{M_j(\eta_<)} d\eta_j  \frac{\Pe}{1 + t \eta_j^2} .
\]
Now by inequality (4.) of  \cite{CombesH-94b} we have
\be
\label{spav}
\int  d\eta_j   \frac{\Pe}{1 + t \eta_j^2}  \le |I| .
\ee
Denote by $ A_>= \{ A_{lk}\}_{l,k \in \L_>}, A_<= \{ A_{lk}\}_{l,k
\in \L_<}$ ''blocks'' of the linear map $A$. From Lemmata \ref{Oberwolfach} and \ref{vol}
below we infer
\[
\vol (M_>^+(\eta_<)) = | \det A_> | \, \omega^{|\L_>|} \, \sum_{n
\in \L_> \cup \{ j\} } |B_{nj}| .
\]
Since $\vol (M_<) = |\det A_< | \, \omega_+^{|\L_<|}$ and Lemma
\ref{trilem} tells us
\[
\det A = A_{jj} \, \det A_< \, \det A_>
\]
we arrive at \bea \int_{\RR^L} d \eta \, k (\eta) \,    {\cal P}
\, (\eta ) & \le & |\det B| \, \omega_+^{-L} \,  ( 1 + t \NO A
\ON_1^2 \omega_+^2) |\det A| \, \omega_+^{|\L_>|+ |\L_<|} \,
\sum_{n \in \L_> \cup \{ j\} } |B_{nj}| \, |I|
\\
& \le & \omega_+^{-1} \, (1+t \NO A \ON_1^2 \omega_+^2) \, \NO B
\ON_1 \, | I| . \eea Taking the limit $t \searrow 0$ we get
\[
\int_{\RR^L} d \eta \, k (\eta) \,    {\cal P} \, (\eta ) \le
\frac{\omega_+}{1-a^*} \, |I|
\]
which proves the proposition for the  case $w = \chi_0$.

Now consider general $w$. We have $ V_{B\eta} = \sum_{j \in \tL}
\eta_j w (\cdot - j ) $  on $\L$ and the spectral averaging
applies as in inequality (\ref{spav}). By independence of the
coupling constants  $\omega_k, k \in \ZZ^d$ we have
\[
\EE \la f , \cj P_\bullet^l (I) \cj f \ra \le \EE \l [
\int_{\RR^L} d \eta \, k (\eta) \,    {\cal P} \, (\eta )       \r
]
\]
and now the proof proceed as in the special case $w = \chi_0$.
\end{pro}
\begin{rem}
\label{Gammacube}
Since $a_j$ may be $0$ for  a $j \in \Gamma$ we can  assume  (by
enlargement) that $\Gamma$ is a discrete cube. It follows that
$\Lambda^+$ is a cube, too. If $\Gamma := \{ \gamma \in \ZZ^d | \,
\gamma_i \in [0 , g] \,  \forall i = 1, \dots , d\}$ and $\Lambda
=\Lambda_l, l \in \NN$, then $\Lp = \{ k \in \ZZ^d | \, k_i \in
[-g, l] \,  \forall i = 1, \dots , d\}$.
\end{rem}

The following lemma is trivial in the case $\ZZ^d = \ZZ$. In the
higher dimensional case it depends on the definition of the
relation ''$\succ $''.
\begin{lem}
\label{trilem}
\begin{enumerate}[(1)]
\item
Assume w.l.o.g.~ that $\Lp$ is a discrete cube, cf.~Remark \ref{Gammacube}. Let $L = \# \Lp$ and $A : \RR^L \to \RR^L, (A\omega)_j := \sum_{k \in \Lp} A_{jk} \omega_k$ be a linear map  as before such that for all $j,k \in \Lp$
\beq \label{lt} A_{jk}\neq 0 \Rightarrow  j \succ k
\\
\label{diag1} A_{jj} = 1  . \eeq
Then $A$ is invertible and the coefficients of $A^{-1} = B = \{ B_{jk} \}_{j,k \in \Lp} $ satisfy (\ref{lt}) and (\ref{diag1}) for all $j,k \in \Lp$.\\
\item
 Let $O \subset \ZZ^d$ be finite and $A : \RR^{|O|} \to \RR^{|O|} $ be given by $ (A\omega)_j := \sum_{k \in O} A_{jk} \omega_k$ with (\ref{lt}) for all $j,k \in O$. Then
\be
\det A = \prod_{j \in O} A_{jj} .
\ee
\end{enumerate}
\end{lem}
\begin{pro}
(1) By part (2) $\det A =\prod_{j \in \Lp} A_{jj} =1$ and $A^{-1}$
exists. We prove by induction over $j \in \Lp$ \be \label{IA}
B_{jj}=1 \ \text{ and } B_{jk} =0 , \ \forall \, k \in \Lp, j \not
\succ k \ee for all $j \in \Lp$. 
Without loss of generality we may assume by
translation $ \Lp = [0, \lambda]^d \cap \ZZ^d$. The induction
anchor is: \[ j =0, k\succ 0: \delta_{0,k}= \sum_{l \in \Lp}
A_{0l} B_{lk} = A_{00} B_{0,k} = B_{0,k}.
\]
Induction step: Let $ m \in \Lp$. If (\ref{IA}) is true for all $j
\in \Lp, j \prec m, j \neq m$ then (\ref{IA}) is true also  for $j
= m$. Proof:
\[
\delta_{mk} = \sum_{j \in \Lp, j\prec m} A_{mj} B_{jk} = \sum_{j
\in \Lp, k \prec j\prec m, j \neq m} A_{mj} B_{jk} + A_{mm}
B_{mk}= B_{mk}
\]
for $k \not \prec i$ or $k =i$.

(2) Let $\Pi_o$ denote the permutation group of $O$. Since $\det A=
\sum_{\pi \in \Pi_O} \prod_{k \in O} A_{k \pi(k)}$ it suffices to
show $ \prod_{k \in O} A_{k \pi(k)}=0$ for all $\pi \neq \Id_O$. For
$\pi \neq \Id_O$ there exists a $ k \in O$ such that $\pi(k) \neq
k$. We claim that there exists a $n \in \NN$ such that
\[
\Lp \ni j := \pi^{n-1}(k) \not \succ \pi^n(k)= \pi (j).
\]
This implies $A_{j \pi(j)} =0 $ and we are finished. To prove the
claim assume $\pi^{n-1} (k) \succ \pi^n(k)$ for all $n \in \NN$.
$\pi(k) \neq k$ implies $\pi^{n-1} (k) \neq \pi^n(k)$ for all $n
\in \NN$. Since $O$ is finite there exist $n,m \in \NN$ such that
$ \pi^n(k) = \pi^{n+m}(k) $. Thus for any $i \in \{1,\dots, d\}$
\[
\pi^n(k)_i \ge  \pi^{n+1} (k)_i \ge  \dots \ge \pi^{n+m} (k)_i =
\ge  \pi^{n} (k)_i \quad \Rightarrow    \pi^{n} (k)_i =  \pi^{n+1}
(k)_i .
\]
Therefore $\pi^{n} (k) =  \pi^{n+1} (k)$ which is a contradiction.
\end{pro}

\begin{lem}
\label{Oberwolfach}
Let $t \in \RR^n$ and
\[
S = \bigcup_{s \in [0,\omega_+]} \{x \in \RR^n| \, x \in
[0,\omega_+]^n + s t  \} .
\]
Then \be \label{diagdet} \vol(S) = \l ( 1 + \sum_{i=1}^n |t_i| \r)
\omega_+^n . \ee
\end{lem}
\begin{pro}
For each $i \in \{1, \dots, n\}$ define the linear map $\tT_i :
\RR^n \to \RR^n$ by
\[
\tT_i (e_l) = e_l \ \text{ for $l \neq i $ and } \tT_i (e_i) =t .
\]
Then $\det \tT_i = t_i$. Define an invertible, affine map $T_i  :
\RR^n \to \RR^n$ by $T_i(x)= \tT_i (x) + \omega_+ e_i$. Set \bea Q
& := & \{  x \in \RR^n | \, x_i  \in [0,\omega_+] \ \forall \ i =
1,\dots, n \}
\\
K_i & := & \{  x \in Q \  | \, x_i = \omega_+ \} , \ \forall \ i =
1,\dots, n
\\
S_i & := & \{  x \in \RR^n | \, x= y  + s \, t, \, s  \in
[0,\omega_+], y \in K_i \} \, . \eea Then we have up to sets of
measure zero the disjoint union
\[
S = Q \, \cup \, \bigcup_{ i = 1}^n S_i .
\]
We prove $ S_i = T_i (Q) $ for all $i = 1, \dots, n$. Since
$T_i(Q) = \tT_i (Q) + \omega_+ e_i$ the claim is equivalent to $
S_i - \omega_+ e_i = \tT_i (Q) $. Now $ y \in K-i$ is equivalent
to $ y_i = \omega_+$ and $ y_l \in [0,\omega_+]$ for all $ l \neq
i$. Thus $ K_i -\omega_+ e_i = \{ x \in Q| \, x_i =0\} =: K^i$. It
follows \bea S_i -\omega_+ e_i & =& \{x| \, x = y - \omega_+ e_i +
s t, y \in K_i, s \in [0,\omega_+] \}
\\
& =& \{x| \, x = z + s t, y \in K^i, s \in [0,\omega_+] \}
\\
& = &\{x| \, x = \sum_{l=1, l \neq i}^n z_l e_l + s t, y \in K^i,
s, z_l \in [0,\omega_+] , \, \forall l =1, \dots, n, l \neq i  \}
\\
& = &\tT_i(Q) . \eea Thus $ S = Q \, \cup \, \bigcup_{ i = 1}^n
T_i (Q)$  and
\[
\vol(s) = \vol(Q) +  \sum_{i=1}^n \, |\det\tT_i| \, \vol(Q) = \l [
1 + \sum_{i=1}^n \, |t_i|\r ] \vol (Q).
\]
\end{pro}
\begin{lem}
\label{vol}
\begin{enumerate}[(1)]
\item
\[
M_>^+ (\eta_<) = A_> (\Xi - \xi b_j) + A_> \l [ \bigcup_{s \in [0,
\omega_+]} ([0, \omega_+]^{|\L_>|} - s b_j)  \r ]
\]
\item
\[
\vol  [M_>^+ (\eta_<) ]= |\det A_>| \, \sum_{n \in \L_> \cup \{j \} } |B_{nj}| \, \omega_+^{|\L_>|}
\]
\end{enumerate}
\end{lem}
\begin{pro}
\bea M_>^+ (\eta_<) & = & \bigcup_{r \in [\xi, \xi + \omega_+]} \{
\eta_> | \, \sum_{l \in \L_>} B_{nl} \eta_l \in  [0, \omega_+] - r
B_{nj} + \Xi \forall n \in \L_> \}
\\
& = & \bigcup_{r \in [\xi, \xi + \omega_+]} \{ \eta_> | \, \eta_>
\in A_> ( [0, \omega_+]^{|\L_>|} ) -r A_> b_j + A_> \Xi  \}
\\
& = &  \dots \eea where we used property (\ref{diagdet}) for the
inversion of the bloc matrix $A_>$. \bea \dots & = & \bigcup_{s
\in [0, \omega_+]} \{  A_> ( [0, \omega_+]^{|\L_>|} ) -s A_> b_j
-\xi A_> b_j + A_> \Xi  \}
\\
& = & A_> ( \Xi - \xi b_j) + \bigcup_{s \in [0, \omega_+]} \l [
A_>  ( [0, \omega_+]^{|\L_>|} - sB_j) \r ]
\\
& = & A_> ( \Xi - \xi b_j) +A_>  \l [  \bigcup_{s \in [0,
\omega_+]} ( [0, \omega_+]^{|\L_>|} - sb_j) \r ].
\eea
This proves the firs claim and
\[
\vol  [M_>^+ (\eta_<) ]
=  |\det A_>| \ \vol  \l [  \bigcup_{s \in [0, \omega_+]} ( [0, \omega_+]^{|\L_>|} - sb_j) \r ].
\]
Lemma \ref{Oberwolfach} and $\l ( 1 + \sum_{n \in \L_> } |B_{nj}|  \r )
= \l ( \sum_{n \in \L_> \cup \{j \} } |B_{nj}|  \r) $ prove the second assertion.
\end{pro}
\begin{rem}
Lemma \ref{trilem} and thus Proposition  \ref{tech} holds true if $\Gamma \subset \{ k
\in \ZZ^d| \, k \prec 0\}$ or if $\Gamma$ is a subset of some other $d$-dimensional
''quadrant''. Probably we can can allow $ \Gamma$ to be a larger set.

Consider the relation on $\RR^d$
\bea
&& k  \succ 0  \Leftrightarrow
\\
&&k_1 \ge 0 \text{ and for all $i = 2, \dots, d$ we have: $  k_i\ge  0 $ if $ k_\nu =0  \, \forall \, \nu = 1, \dots, i-1$ }
\eea
For this relation and $\Gamma \subset \{ k \in \ZZ^d | \, k \succ 0\}$ the proof should work, too. The reason ist that $ k \succ 0 $ and $ -k \succ 0$ imply $k =0$.
\end{rem}
\section{Discussion of recent results on Wegner estimates for indefinite potentials}
\subsection*{Results from \cite{Veselic-2000b} concerning differentiable densities}
Assume the hypotheses of Theorem \ref{theorem1} up to following changes
\begin{itemize}
\item
the ''support'' $\Gamma$ of the convolution vector is an arbitrary finite subset of $\ZZ^d$.
\item
the single site measure $\mu$ has a density $g\in W^{1,1}(\RR)$.
\end{itemize}
Denote as in Section \ref{proof} by $B$ the inverse of the matrix $A := \{ A_{j,k} \}_{j,k\in \Lp}, \ A_{j,k}:=
a_{j-k}, \forall \, j,k\in \Lp $. In \cite{Veselic-2000b} the following Wegner estimate is proven:
\begin{thm}
We have for all $E \in \RR$
\begin{equation}
\label{resultat2}
\EE \l [ \Tr P_\omega^l( \l [E -\epsilon,E \r ]) \r ]
\le \mbox{const } \NO B\ON  \epsilon \, l^d
\le \mbox{const } \frac{\epsilon \, l^d}{1-a^*}  , \quad \forall \, \epsilon \ge 0.
\end{equation}
The constant depends on $E$ but not on $\epsilon$.
\end{thm}

\begin{rem}
In \cite{Veselic-2000b} the geometric series is used to deduce $\NO B\ON \le \frac{1}{1-a^*}$ and thereby the
second inequality in \eqref{resultat2}. Alternatively one can use criteria formulated in terms of the {\em symbol} $S_A$ of the (block) Toeplitz matrix $A$
which alow one to control the behaviour of the eigenvalue $\nu(l) $ of $A=A_\Lp, \Lp= \Lp_l$ closest to $0$ as $\L = \L_l$ tends to the whole space $\RR^d$. If we can show that $ |\nu(l)| $ tends to zero not faster than a a inverse power of  $l$ we have by  \eqref{resultat2} a Wegner estimate which can be used for the multi scale analysis.
\end{rem}
We discuss first the one dimensional case $d=1$. There is a series of papers by S.~Serra where the assumes that the symbol
\bea
S_A(\theta) = \sum_{k\in \ZZ} a_k e^{i k \theta}, \ \theta \in [-\pi , \pi] , \ i = \sqrt{-1}
\eea
is a real function assuming non-negative values. This corresponds to the case that the matrix $A$ is selfadjoint and non negative. In \cite{Serra-1998a}  it is proven  that if  $S_A$ has one single zero of order $n$ then $|\nu(l)|^{-1}= {\cal O }(l^n) $. This means for our situation that we obtain a Wegner estimate with corresponding volume dependence
\bea
\EE \l [ \Tr P_\omega^l( \l [E -\epsilon,E \r ]) \r ]
\le \mbox{const } \epsilon \, l^{n+1}  , \quad \forall \, \epsilon \ge 0.
\eea
In the article \cite{Serra-1996} Serra considers a similar situation, but now $S_A$ is allowed to have several minima, and finally in \cite{Serra-1994,Serra-1998b} the block-Toeplitz case is considered. Similar results are obtained by Böttcher and Grudsky in \cite{BoettcherG-1998}.

\subsection*{Results from \cite{HislopK-2001}}
In the paper \cite{HislopK-2001} Hislop and Klopp prove a Wegner estimate for indefinite alloy type models. Their proof does not require any condition on the form of the single site potential $u$ as we do in Assumption \ref{asp} \eqref{step}. Their result is not sufficient to imply the existence of the density of states.
  The results  in  \cite{HislopK-2001}  are restricted to energy regions which do  not belong to the spectrum of the unperturbed operator $H_0$. The reason is that  they make use of a different type of restriction of the operator  $H_\omega$ to the finite cube $\Lambda=\Lambda_l$. Namely, the operator associated to  $\Lambda$ is $H_\omega^l := H_0 +V_\omega^l$ where $V_\omega^l(x)$ stands for $\sum_{k \in \tL} \omega_k u(x-k)$.

Hislop and Klopp assume that the single site potential $u$  is continuous and compactly supported and has no zero in a neighbourhood of $0 \in \RR^d$. The density $g$ of the single site distribution has to be from $L_0^\infty(\RR)$ and be locally absolutely continuous.
\begin{thm}[\cite{HislopK-2001}]
For any $q<1$ and $E_0 <\inf \sigma(H_0)$ there exists $C \in ]0,\infty[$ such that for all $\epsilon > 0$
with $ E_0 +\epsilon < \inf\sigma(H_0)$ one has
\bea
\PP\{\omega| \, \sigma(H_\omega^l) \, \cap \, [E_0 -\epsilon, E_0 +\epsilon] \neq \emptyset \}
\le C\epsilon^q |\Lambda|
\eea
where the constant depends only on $d,q$ and the distance between $E_0$ and the unperturbed spectrum $\sigma(H_0)$.
\end{thm}

The Wegner estimate is also true at internal spectral edges (away from the unperturbed spectrum $\sigma(H_0)$) if one works in the weak coupling regime. This means that the considered operator is $H_0 + \lambda V_\omega$ with $|\lambda|$ sufficiently small.
Moreover their proof applies also to the case where the single site potentials have different shapes instead of being the translates $u_k = u (\cdot-k)$ of a single function $u$ and for certain families of correlated coupling constants $\omega_k, k\in \ZZ^d$ and also to certain random operators where the randomness enters via an multiplicative perturbation. For details see \cite{HislopK-2001}.

\begin{rem}[Birman-Schwinger Principle]
In \cite{HislopK-2001} actually an auxiliary operator of Birman-Schwinger type is introduced and  the behaviour of its eigenvalues is analyzed, rather than the one of $H_\omega^l$ itself.

Namely, for an $ E \in (\RR \setminus \sigma(H_\omega^l )) \, \cap \, ]-\infty, \inf \sigma (H_0)[ \,$ one defines the selfadjoint and compact operator
\bea
\Gamma_\omega^l(E):= (H_0-E )^{-1/2} V_\omega^l(H_0-E )^{-1/2}
\eea
and writes now the resolvent of $H_\omega^l$ as
\bea
(H_\omega^l-E)^{-1/2} = (H_0-E )^{-1/2} [ 1+ \Gamma_\omega^l(E)]^{-1} (H_0-E )^{-1/2}
\eea
whose norm is bounded by
\bea
\delta \, \| [1+ \Gamma_\omega^l(E)]^{-1}\|, \quad \delta  :=[\inf \sigma (H_0)-E]^{-1} .
\eea
Having this in mind one can reformulate the Wegner estimate as follows
\begin{align}
\nn
\PP \{ \omega | \, d(\sigma(H_\omega^l), E) < \epsilon\}
& =
\PP \{ \omega | \, \|(H_\omega^l-E)^{-1}\| > \epsilon^{-1}\}
\\
\label{BS}
& \le
\PP \{ \omega | \, \|[1+ \Gamma_\omega^l(E)]^{-1}\| > \delta/\epsilon\}
\\
\nn
& =
\PP \{ \omega | \, d(\sigma[\Gamma_\omega^l(E)],-1) < \delta/\epsilon\}
\end{align}
use the \v Ceby\v sev inequality
\begin{align}
\PP \{ \omega | \, d(\sigma[\Gamma_\omega^l(E)],-1) < \delta/\epsilon\}
&\le
\EE \{\Tr\chi_{]-1-\delta/\epsilon,-1+\delta/\epsilon[} ( \Gamma_\omega^l(E)) \}
\end{align}
and proceed with the spectral analysis of the Birman-Schwinger operator $\Gamma_\omega^l(E)$, cf. \cite{Klopp-95a,HislopK-2001}.
\end{rem}

\small
\bibliographystyle{alpha}

\begin{thebibliography}{FHLM97}

\bibitem[BG98]{BoettcherG-1998}
A. B{\"o}ttcher and S.~M. Grudsky.
\newblock On the condition numbers of large semi-definite {T}oeplitz matrices.
\newblock {\em Linear Algebra Appl.}, 279(1-3):285--301, 1998.

\bibitem[CH94]{CombesH-94b}
J.-M. Combes and P.D. Hislop.
\newblock Localization for some continuous, random {Hamiltionians} in
  d-dimensions.
\newblock {\em J. Funct. Anal.}, 124:149--180, 1994.

\bibitem[CHT99]{CombesHT-1999}
J.~M. Combes, P.~D. Hislop, and A.~Tip.
\newblock Band edge localization and the density of states for acoustic and
  electromagnetic waves in random media.
\newblock {\em Ann. Inst. H. Poincar\'e Phys. Th\'eor.}, 70(4):381--428, 1999.

\bibitem[CL90]{CarmonaL-90}
R.~Carmona and J.~Lacroix.
\newblock {\em Spectral Theory of Random {Schr\"odinger} Operators}.
\newblock Birkh\"auser, Boston, 1990.

\bibitem[CT73]{CombesT-73}
J.M. Combes and L.~Thomas.
\newblock Asymptotic behaviour of eigenfunctions for multiparticle
  {Schr\"odinger} operators.
\newblock {\em Commun. Math. Phys.}, 34:251--270, 1973.

\bibitem[FHLM97]{FischerHLM-1997}
W.~Fischer, T.~Hupfer, H.~Leschke, and P.~M{\"u}ller.
\newblock Existence of the density of states for multi-dimensional continuum
  {S}chr\"odinger operators with {G}aussian random potentials.
\newblock {\em Comm. Math. Phys.}, 190(1):133--141, 1997.

\bibitem[FK96]{FigotinK-1996}
A.~Figotin and A.~Klein.
\newblock Localization of classical waves. {I}. {A}coustic waves.
\newblock {\em Comm. Math. Phys.}, 180(2):439--482, 1996.

\bibitem[FK97]{FigotinK-1997a}
A.~Figotin and A.~Klein.
\newblock Localization of classical waves. {I}{I}. {E}lectromagnetic waves.
\newblock {\em Comm. Math. Phys.}, 184(2):411--441, 1997.

\bibitem[FLM00]{FischerLM-2000}
W.~Fischer, H.~Leschke, and P.~M{\"u}ller.
\newblock Spectral localization by {G}aussian random potentials in
  multi-dimensional continuous space.
\newblock {\em J. Statist. Phys.}, 101(5-6):935--985, 2000.

\bibitem[FS83]{FroehlichS-83}
J.~Fr\"ohlich and T.~Spencer.
\newblock Absence of diffusion in the {Anderson} tight binding model for large
  disorder or low energy.
\newblock {\em Commun. Math. Phys.}, 88:151--184, 1983.

\bibitem[GK01a]{GerminetK-2001a}
F.~Germinet and A.~Klein.
\newblock Bootstrap multiscale analysis and localization in random media.
\newblock {\em Comm. Math. Phys.}, 222(2):415--448, 2001.

\bibitem[GK01b]{GerminetK-2001b}
F.~Germinet and A.~Klein.
\newblock A characterization of the {Anderson} metal-insulator transport
  transition.
\newblock Preprint, www.ma.utexas.edu/mp\_arc/, 2001.

\bibitem[HK01]{HislopK-2001}
P.~D. Hislop and F.~Klopp.
\newblock The integrated density of states for some random operators with
  nonsign definite potentials.
\newblock www.ma.utexas.edu/mp\_arc, preprint no. 01-139, 2001.

\bibitem[How87]{Howland-87a}
J.~S. Howland.
\newblock Perturbation theory of dense point spectra.
\newblock {\em J. Func. Anal.}, 74:52--80, 1987.

\bibitem[HS96]{HislopS-96}
P.~D. Hislop and I.M. Sigal.
\newblock {\em Introduction to spectral theory: with Applications to
  {Schr\"odinger} Operators}.
\newblock Springer, New York, 1996.

\bibitem[Kir89]{Kirsch-89a}
W.~Kirsch.
\newblock Random {Schr\"odinger} operators.
\newblock In H.~Holden and A.~Jensen, editors, {\em {Schr\"odinger} Operators},
  Lecture Notes in Physics, {\bf 345}, Berlin, 1989. Springer.

\bibitem[Klo95]{Klopp-95a}
F.~Klopp.
\newblock Localization for some continuous random {Schr\"odinger} operators.
\newblock {\em Commun. Math. Phys.}, 167:553--569, 1995.

\bibitem[Klo99]{Klopp-1999}
F.~Klopp.
\newblock Internal {L}ifshits tails for random perturbations of periodic
  {S}chr\"odinger operators.
\newblock {\em Duke Math. J.}, 98(2):335--396, 1999.

\bibitem[KM82]{KirschM-82c}
W.~Kirsch and F.~Martinelli.
\newblock On the density of states of {Schr\"odinger} operators with a random
  potential.
\newblock {\em J. Phys. A: Math. Gen.}, 15:2139--2156, 1982.

\bibitem[KS87]{KotaniS-87}
S.~Kotani and B.~Simon.
\newblock Localization in general one-dimensional random systems {II}:
  continuum {Schr\"odinger} operators.
\newblock {\em Commun. Math. Phys.}, 112:103--119, 1987.

\bibitem[KSS98]{KirschSS-1998a}
W.~Kirsch, P.~Stollmann, and G.~Stolz.
\newblock Localization for random perturbations of periodic {S}chr\"odinger
  operators.
\newblock {\em Random Oper. Stochastic Equations}, 6(3):241--268, 1998.
\newblock available at www.ma.utexas.edu/mp\_arc, preprint no. 96-409.

\bibitem[MH84]{MartinelliH-84}
F.~Martinelli and H.~Holden.
\newblock On absence of diffusion near the bottom of the spectrum for a random
  {Schr\"odinger} operator on {$ L^{2}(R^{\nu}) $}.
\newblock {\em Commun. Math. Phys.}, 93:197--217, 1984.

\bibitem[MS85]{MartinelliS-85}
F.~Martinelli and E.~Scoppola.
\newblock Remark on the absence of the absolutely continuous spectrum for
  d-dimensional {Schr\"odinger} operator with random potential for large
  disorder or low energy.
\newblock {\em Commun. Math. Phys.}, 97:465--471, 85.

\bibitem[PF92]{PasturF-92}
L.~A. Pastur and A.~L. Figotin.
\newblock {\em Spectra of Random and Almost-Periodic Operators}.
\newblock Springer Verlag, Berlin, 1992.

\bibitem[RS78]{ReedS-78}
M.~Reed and B.~Simon.
\newblock {\em Methods of Modern Mathematical Physics {IV}, Analysis of
  Operators}.
\newblock Academic Press, San Diego, 1978.

\bibitem[Ser94]{Serra-1994}
S.~Serra.
\newblock Preconditioning strategies for asymptotically ill-conditioned
  {Toeplitz} matrices.
\newblock {\em BIT}, 34:579--593, 1994.

\bibitem[Ser96]{Serra-1996}
S.~Serra.
\newblock On the extreme spectral properties of {T}oeplitz matrices generated
  by ${L}\sp 1$ functions with several minima/maxima.
\newblock {\em BIT}, 36(1):135--142, 1996.

\bibitem[Ser98a]{Serra-1998a}
S.~Serra.
\newblock On the extreme eigenvalues of {H}ermitian (block) {T}oeplitz
  matrices.
\newblock {\em Linear Algebra Appl.}, 270:109--129, 1998.

\bibitem[Ser98b]{Serra-1998b}
S. Serra.
\newblock Asymptotic results on the spectra of block {Toeplitz} preconditioned
  matrices.
\newblock {\em SIAM J. Matrix Anal.Appl.}, 20:31--44, 1998.

\bibitem[Sto98]{Stollmann-1998}
P.~Stollmann.
\newblock Localization for random perturbations of anisotropic periodic media.
\newblock {\em Israel J. Math.}, 107:125--139, 1998.

\bibitem[Sto01]{Stollmann-2001}
P.~Stollmann.
\newblock {\em Caught by disorder: A Course on Bound States in Random Media},
  volume~20 of {\em Progress in Mathematical Physics}.
\newblock Birkh\"auser, July 2001.

\bibitem[SW86]{SimonW-86}
B.~Simon and T.~Wolff.
\newblock Singular continuous spectrum under rank one perturbations and
  localization for random {Hamiltonians}.
\newblock {\em Comm. Pure Appl. Math.}, 39:75--90, 1986.

\bibitem[Ves98]{Veselic-1998}
I.~Veseli\'c.
\newblock Localisation for random perturbations of periodic {Schr\"odinger}
  operators with regular {Floquet} eigenvalues.
\newblock  to appear in {\em Ann.~Henri Poincar\'e}, available at
  www.ma.utexas.edu/mp\_arc preprint no. 98-569, 1998.

\bibitem[Ves00]{Veselic-2000b}
I.~Veseli\'c.
\newblock Wegner estimate for some indefinite {Anderson}-type {Schr\"odinger}
  operators with differentiable densities, to appear in {\em Lett Math. Phys.}
\newblock preprint 2000, www.ma.utexas.edu/mp\_arc/.

\bibitem[Ves01]{Veselic-2001}
I.~Veseli\'c.
\newblock {\em Indefinite Probleme bei der Anderson-Lokalisierung}.
\newblock {Ph.D} thesis, Ruhr-Universit\"at Bochum, 44780 Bochum, January 2001.
\newblock http://www-brs.ub.ruhr-uni-bochum.de/netahtml/HSS/Diss/VeselicIvan/.

\bibitem[Weg81]{Wegner-81}
F.~Wegner.
\newblock Bounds on the {DOS} in disordered systems.
\newblock {\em Z. Phys. B}, 44:9--15, 1981.

\bibitem[Zen99]{Zenk-1999}
H.~Zenk.
\newblock Anderson localization for a multidimensional model including long
  range potentials and displacements.
\newblock {Preprint-Reihe} des {Fachbereichs Mathematik} at the {Johannes}
  {Guteberg}-Universit\"at {Mainz}, 1999.

\end{thebibliography}
\def\cprime{$'$}

\end{document}